\documentclass{article}
\usepackage{mathrsfs,amsmath,graphicx}

\usepackage{authblk}

\title{Numerical correction of anti-symmetric aberrations in single HRTEM images of weakly scattering 2D-objects}

\author[1,*]{Ossi Lehtinen}
\author[1]{Dorin Geiger}
\author[1]{Zhongbo Lee}
\author[2]{Michael Brian Whitwick}
\author[2]{Ming-Wei Chen}
\author[2]{Andras Kis}
\author[1]{Ute Kaiser}

\affil[1]{Central Facility for Electron Microscopy, Group of Electron Microscopy of Materials
Science, University of Ulm, 89081 Ulm, Germany}
\affil[2]{Electrical Engineering Institute, Ecole Polytechnique Federale de Lausanne (EPFL), CH-1015 Lausanne, Switzerland}
\affil[*]{email: ossi.lehtinen@gmail.com}

\begin{document}

\maketitle

\begin{abstract}
Here, we present a numerical post-processing method for removing the effect of anti-symmetric residual aberrations in high-resolution transmission electron microscopy (HRTEM) images of weakly scattering 2D-objects. The method is based on applying the same aberrations with the opposite phase to the Fourier transform of the recorded image intensity and subsequently inverting the Fourier transform. We present the theoretical justification of the method and its verification based on simulated images in the case of low-order anti-symmetric aberrations. Ultimately the method is applied to experimental hardware aberration-corrected HRTEM images of single-layer graphene and MoSe$_2$ resulting in images with strongly reduced residual low-order aberrations, and consequently improved interpretability. Alternatively, this method can be used to estimate by trial and error the residual anti-symmetric aberrations in HRTEM images of weakly scattering objects.
\end{abstract}

\section{Introduction}

Thanks to the practical realization of hardware aberration correction (AC), the point resolution of high-resolution transmission electron microscopy (HRTEM) is not limited by the positive spherical aberration ($C_s$) any longer~\cite{haider1998,scherzer1936,uhlemann1998}, and materials can now be routinely imaged at atomic resoultion with state-of-the-art instruments even at low acceleration voltages~\cite{meyer2008,krivanek2010}. Once the strong $C_s$ contribution is removed, other aberrations such as three-fold astigmatism ($A_2$) and coma ($B_2$) become measurable and can be corrected~\cite{uhlemann1998} down to a limit imposed by the measurement accuracy and the adjustment precision of the correcting elements. With advanced techniques, such as electron holography~\cite{gabor1948,lichte1986electron,geiger2008,Linck201377} and exit wave reconstruction from a focal series~\cite{saxton1994focus,op1996wave,kirkland1995super,coene1996maximum,kirkland1997multiple} direct information on the specimen, not obscured by the inevitable residual aberrations, can be acquired. However, these techniques require special arrangements/procedures during the acquisition of the images, and thus cannot be applied to a single HRTEM image {\em post-situ}.

The wave aberrations distort the information transferred from the specimen to the imaging device, such as a CCD camera, making interpretation of the acquired images difficult if not sometimes impossible~\cite{krivanek1995}. In a corrected microscope this problem is to a large extent overcome, as is evident from the nearly aberration-free images of today. In the context of this article, the word {\em nearly} should be emphasized, however. The aberration coefficients can be tuned down to zero only with some accuracy. Thus, in practice residual aberrations are always present even in corrected instruments~\cite{meyer2011,biskupek2012}. Moreover, the instruments tend to drift away from the corrected state over time, and one can expect stronger residual aberrations as time passes after tuning the corrector~\cite{Barthel20136}. Residual aberrations can be also of benefit, like in the case of $C_s$, as non-zero $C_s$ is required for optimal phase contrast transfer~\cite{uhlemann1998}. 

\begin{figure}
\includegraphics[width=\textwidth]{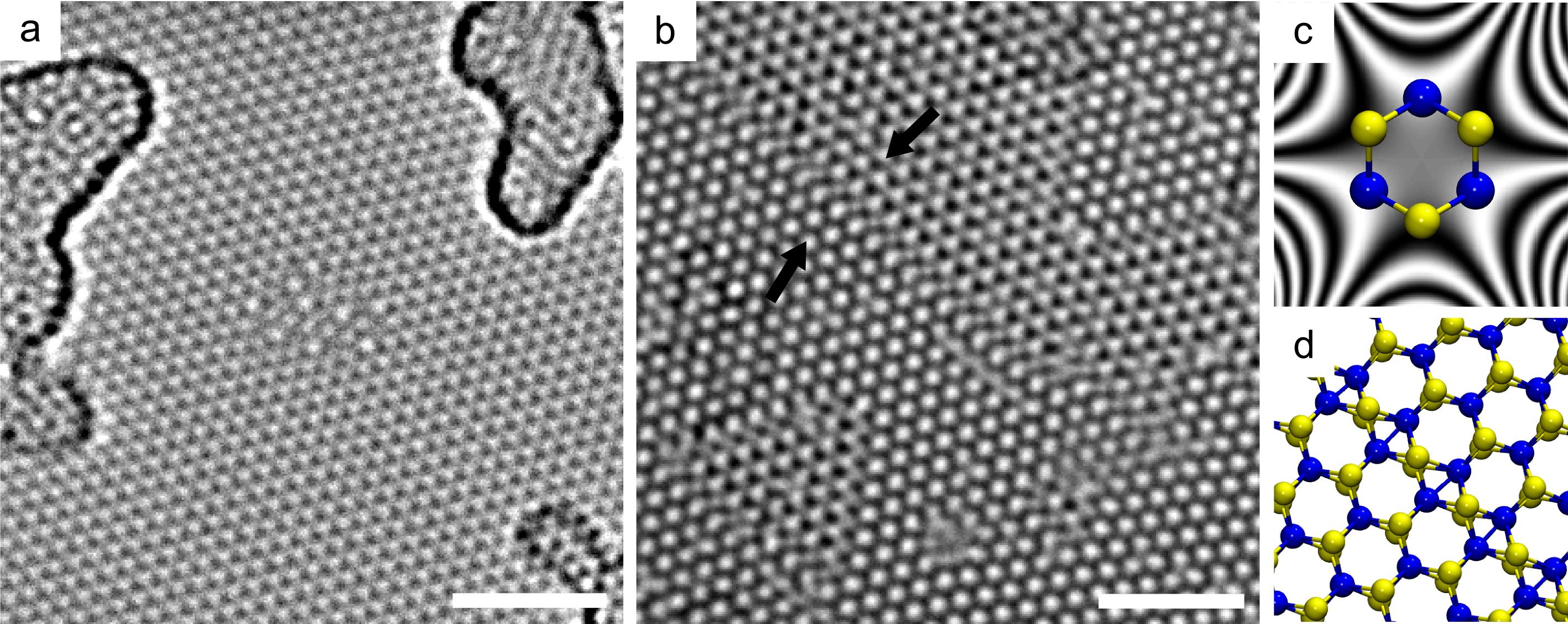}
\caption{{\bf Experimental AC-HRTEM images of graphene and MoSe$_2$ suffering from residual aberrations.} {\bf a}: A defect structure in single-layer graphene, where the graphene lattice does not appear hexagonal due to residual aberrations. This effect can be attributed to residual three-fold astigmatism ($A_2$), which induces a contrast difference between the sublattices of graphene, causing the lattice image to deviate from the hexagonal symmetric pattern expected for graphene. {\bf b}: Mirror-twin-boundaries in single-layer MoSe$_2$. Due to residual aberrations, the lattice has a completely different appearance on the opposite sides of the boundaries. Compare, for example, the two areas marked by the arrows. Also this effect can be explained by alignment of the MoSe$_2$ lattice with residual $A_2$ (panel {\bf c}) and by the swapping of the Mo and dual Se sites in the three-fold symmetric lattice when crossing the boundary (see panel {\bf d} for the structure, where the dark atoms represent Mo and light atoms Se, with two Se atoms always residing on top of each other). The reversal of the lattice results in the accentuation of different sublattices in these regions. The scale bars are 2~nm.}
\label{residualA2}
\end{figure}

On the other hand, residual aberrations such as $A_2$ and $B_2$, which can be present in the order of 100~nm after correction, can lead to undesired artefacts in the images, and the effect of these can be anything from a minor nuisance to complete misinterpretation of the image. Examples of such effects are presented in Figure~\ref{residualA2}. In the first panel, an experimental aberration-corrected HRTEM (AC-HRTEM) image of a point defect in single-layer graphene can be seen. The presence of residual aberrations in the image is clear from the strongly non-hexagonal appearance of the graphene lattice. This effect can be attributed to the three-fold astigmatism $A_2$, which has three axes of symmetry in the image plane (see Figure~\ref{residualA2}c for a visualization): Depending on the relative orientation of $A_2$ to the graphene lattice, one of the two sublattices of graphene can produce a higher contrast than the other, resulting in the observed non-hexagonal pattern. The atomic structure of the defect can still be deduced from the experimental image, but the image is far from optimal. 

In the second panel, an experimental AC-HRTEM image of so-called mirror-twin-boundaries in a single-layer MoSe$_2$ sample can be seen. In an aberration-free image the Se sites produce higher contrast as compared to the Mo sites, as there are always two Se (Z=34) atoms at the same location as opposed to single Mo (Z=42) atoms. However, in the recorded image the lattice has a completely different appearance at the different sides of the linear boundaries (compare, for example, the areas denoted by the two arrows). On one side the second sublattice has a much higher contrast, but on the other side the sublattices have almost equal contrast, resulting in the hexagonal appearance of the lattice. This effect can again be attributed to residual $A_2$, which is oriented matching the sublattices of the crystal with trigonal symmetry (see Figure~\ref{residualA2} panel c). On one side of the boundary the Se sites are further enhanced, increasing the contrast difference of the Mo and Se sites. However, when crossing the boundary, the Mo and Se sites are interchanged (see Figure~\ref{residualA2} panel d), and now the $A_2$ has the effect of reducing the contrast difference resulting in nearly equal contrast for the Mo and Se sites. Evaluation of the contrast could lead one to deduce that there are two different materials on the opposite sides of the boundary, which in fact is not the case. We would like to emphasize that these artefacts are not introduced by exceptionally strong residual aberrations. Under our experimental conditions, the $A_2$ coefficient is in the order of 100~nm, which is a typical state of correction, after tuning, in our microscope (FEI Titan 80-300, operated at 80~kV).

As mentioned above, numerous methods for acquiring truly aberration-free images in TEM have been proposed and implemented in practice. The common denominator of these methods is that they aim at recovering the object wave function, which then can be manipulated much more freely and interpreted more directly than a simple intensity image~\cite{geiger2008,Linck201377,kirkland1995super,coene1996maximum,kirkland1997multiple,Thust1996249}. In electron holography~\cite{gabor1948,lichte1986electron}, an iterference pattern between the object wave and a reference wave is generated with the help of a bi-prism, and direct information on the object wave can be recovered numerically from this pattern. The $C_s$ corrector improves the phase-detection limit of electron holography significantly~\cite{geiger2008}. Another group of methods is in-line holography, {\em i.e.}, exit wave reconstruction~\cite{saxton1994focus,op1996wave,kirkland1995super,coene1996maximum,kirkland1997multiple,saxton1988accurate}, where typically a series of images at equally spaced foci is acquired and subsequently the object wave is deduced iteratively. After the 'raw' electron wave is recovered, the influence of any residual aberrations can be corrected~\cite{Thust1996249,geiger2008,Linck201377}. A common drawback of these methods, however, is that they all require special arrangements and/or procedures during acquisition of the images, and hence cannot be applied to a single HRTEM image during post-processing. Such a situation can be quite frustrating if the microscopist has managed to capture an elusive target during imaging, but the resulting image is of poor quality.

Here, we present a method for numerically removing the effects of a specific group of geometric aberrations (the anti-symmetric ones, as discussed later) {\em after} acquisition in single HRTEM images of weakly scattering objects. The requirement of the investigated object to be a weak scatterer is a strong one, and precludes a large number of typically studied materials from the applicability range of the method. However, one important and intensively studied group of materials, that is, the recently discovered class of 2D materials with graphene as the prime example, do approximately satisfy this condition at typically used acceleration voltages such as 80~kV~\cite{Lee201239}, making the method potentially useful in a large number of studies.

In analogy to the phase reconstruction methods (focal series and off-axis holography) where residual aberrations are removed from the wave function after the reconstruction, we remove the contribution of the residual anti-symmetric aberrations using the fast Fourier transform (FFT), but instead of the full wave function, we work on a single image, which is possible in the case of a weakly scattering object, that is, when the linear imaging theory is valid~\cite{tembook}. Because the weakly scattering objects we investigate are 2D-objects like graphene, the tilt angles are usually small and the effect of object tilt will be neglected in the first approximation. We want to make clear, that the effect of the symmetric aberrations, such as defocus or $C_s$ cannot be remedied by our method. This is clear already from the fact that the symmetric aberrations lead to loss of information at certain frequencies (the zero crossing points of the contrast transfer function), which of course cannot be recovered. We find our method to be particularly useful for removing the effects of residual aberrations in distorted AC-HRTEM images of weakly scattering 2D-materials, where information is available on how the image {\em should} look like, as demonstrated with the examples of graphene and MoSe$_2$.

\section{Methods}

First, we present the theoretical background and justification of our method. We follow closely the notation of Ref.~\cite{ishizuka2013phase}.

Assuming coherent illumination and exluding the damping envelopes from the treatment, the wave function $\Psi_z({\bf x})$ at the image plane at a distance $z$  from the object along the optical axis can be written as

\begin{equation}
\Psi_z({\bf x}) = \Psi_0({\bf x}) \otimes p_z({\bf x}) = (1+\phi({\bf x}))\otimes p_z({\bf x}),
\end{equation}

where $\Psi_0({\bf x})$ is the wave at the exit surface, $\phi({\bf x})$ is the scattered wave, $p_z({\bf x})$ is the lens transfer function and $\otimes$ denotes convolution. The image intensity $i_z({\bf x})$ is the square modulus of the wave at the detector plane, and its Fourier transform $I_z({\bf u})$ can be written as

\begin{equation}
\label{eq1}
\begin{split}
I_z({\bf u}) & \equiv  \mathcal{F}\{\Psi_z\cdot\Psi^*_z\}\\
& = \mathcal{F}\{ 1 + \phi\otimes p_z + \phi^*\otimes p^*_z + (\phi\otimes p_z)(\phi^*\otimes p^*_z)\},
\end{split}
\end{equation}

In the case of a weakly scattering object (that is, when the weak phase object approximation is valid), an approximation can be made here~\cite{tembook}. That is, the last term on the second line of the equation can be deemed to be small, which is in effect a transition from the general non-linear imaging theory to the special case of linear imaging theory. Fortunately, 2D-materials such as graphene fulfil this condition at typical acceleration voltages such as 80~kV~\cite{Lee201239} and, as demonstrated by simulations and experiments below, the method can be applied on images of such materials. Equation~2 now becomes

\begin{equation}
I_z({\bf u}) \approx  \delta({\bf u}) + \Phi({\bf u})\cdot e^{-i\chi({\bf u})} + \Phi^*(-{\bf u})\cdot e^{i\chi(-{\bf u})},
\end{equation}

where $\Phi({\bf u})$ denotes the Fourier transform of $\phi({\bf x})$. The Fourier transform of $p_z$ is written in the explicit form $e^{-i\chi({\bf u})}$. The information about the microscope aberrations are encoded in the wave aberration function $\chi({\bf u})$.

Now, $\chi({\bf u})$ can be represented as a power series where each term represents a specific aberration. The terms of the series can be divided into symmetric $\chi_s({\bf u}) = \chi_s(-{\bf u})$ and anti-symmetric $\chi_{as}({\bf u}) = -\chi_{as}(-{\bf u})$ parts. Notably, the defocus and $C_s$ belong to the symmetric group and $A_2$ and $B_2$ to the anti-symmetric group. Thus, $I_z({\bf u})$ can be rewritten as 

\begin{equation}
I_z({\bf u}) = \delta({\bf u}) + \Phi({\bf u})\cdot e^{-i\chi_s({\bf u})}e^{-i\chi_{as}({\bf u})} + \Phi^*(-{\bf u})\cdot e^{i\chi_s({\bf u})}e^{-i\chi_{as}({\bf u})}.
\end{equation}

From this equation it is evident, that the effect of all the {\em anti-symmetric} aberrations can be removed from the image by simply multiplying $I_z({\bf u})$ by $e^{i\chi_{as}({\bf u})}$ and calculating the inverse Fourier transform. That is, a corrected image $i_c({\bf x})$ can be generated by

\begin{equation}
i_{c}({\bf x}) = |\mathcal{F}^{-1}\{ e^{i\chi_{as}({\bf u})}\cdot \mathcal{F}\{i_z({\bf x})\}\}|.
\label{corrector}
\end{equation}

The problem of correcting the residual aberrations is two-fold, however. So far we have shown that, in principe, the anti-symmeric aberration contribution in a HRTEM image of a weakly scattering object can be freely manipulated in post-processing. However, in order to do actual correction, the aberration coefficients and orientations need to be determined. This is a shared challenge with all the methods aiming to correct aberrations in TEM, and of course one is limited by the accuracy at which the aberrations can be measured. Information on the residual aberrations is offered by the corrector based on a series of tilted-beam images~\cite{uhlemann1998,barthel2007}. Other methods for measuring the residual aberrations, {\em e.g.}, from a focus/tilt series~\cite{meyer2004a,meyer2004b} or even a single HRTEM image~\cite{stenkamp1998a} have been presented in the literature. The aberrations could, in priciple, be removed with the corrector once determined, but, {\em e.g.}, due to drift of the corrector state during searching of an area of interest and image acquisition~\cite{Barthel20136} and finite precision of the corrector adjustment, typically residual aberrations in the final images are inevitable. 

Post-acquisition correction by the harware corrector is of course not possible, and one needs to find new strategies for fine-tuning the existing images if necessary. One alternative is to look at a feature of an image ({\em e.g.}, the pristine graphene lattice) and find the coefficients which result in the expected appearance for an aberration-free image by trial and error. Great care should be taken with this approach, however, as the problem can be underdetermined. For example, in our example case of three-fold $A_2$ astigmatism and MoSe$_2$ (see e.g. the illustration in Figure~\ref{residualA2} panel c) the same sublattice imbalance can be introduced by different combinations of amplitude and relative orientation of the $A_2$ and the MoSe$_2$ lattice. A conservative approach in such a situation is to find the orientation where the smallest possible amplitude corrects the visible effect. If the actual orientation of the real residual aberration would be different from this, some residual aberration would remain (the new residual aberration is the vector sum of the real aberration and the correcting aberration), but the situation would still be improved and the aberration would not be 'over corrected'. The situation is further complicated, if multiple dominant residual aberrations are present. However, often the third-order anti-symmetric aberrations, that is $B_2$ and $A_2$ are dominating (assuming defocus and first-order astigmatism are properly tuned). In the special case of 2D-materials, which tend to have a three-fold symmetry, the three-fold symmetric $A_2$ has a distinctive effect on the appearance of the lattice, and thus can be distinguished from the effects of $B_2$. Thus, even if the aberration post-correction is not ideal, a significantly better result can be achieved in suitable situations, as shown below by experiments and simulations.

All the experimental images presented here were acquired using an FEI Titan 80-300 microscope equipped with an image-side hexapole aberration corrector~\cite{haider1998}. The microscope was operated at 80~kV and a reduced extraction voltage of 2~kV, in order to reduce the energy spread of the electron beam. The $C_s$ was corrected to $<$20~$\mu$m and the $A_2$ and $B_2$ were corrected to the order of 100~nm (unless stated otherwise).

The graphene samples were produced by the chemical vapor deposition method (acquired commercially from Graphenea S. A.) and transferred on Quantifoil TEM-grids by the procedure described in Ref.~\cite{algara2014dry}. The defect in Figure~\ref{residualA2}a is produced by carbon deposition, as will be described in detail in a separate paper~\cite{lehtinen2014implantation}. The MoSe$_2$ sample was produced by molecular beam epitaxy and transferred on to Pelco holey silicon nitride TEM grids (a detailed study on this sample will be presented also in a separate article~\cite{lehtinen2014microstructure}).

The image simulations were conducted using the QSTEM software package~\cite{qstem}, using $C_s = $~20~$\mu$m and focus spread of 6~nm. The simulated structure consisted of 149 Mo and 288 Se atoms. and the simulation cell size was 6~nm$\times$6~nm with a sampling of 0.03~\AA/pixel (the panels in Figure~\ref{simulated_correction} are cropped from the top right).

\section{Results}

\begin{figure}
\includegraphics[width=\textwidth]{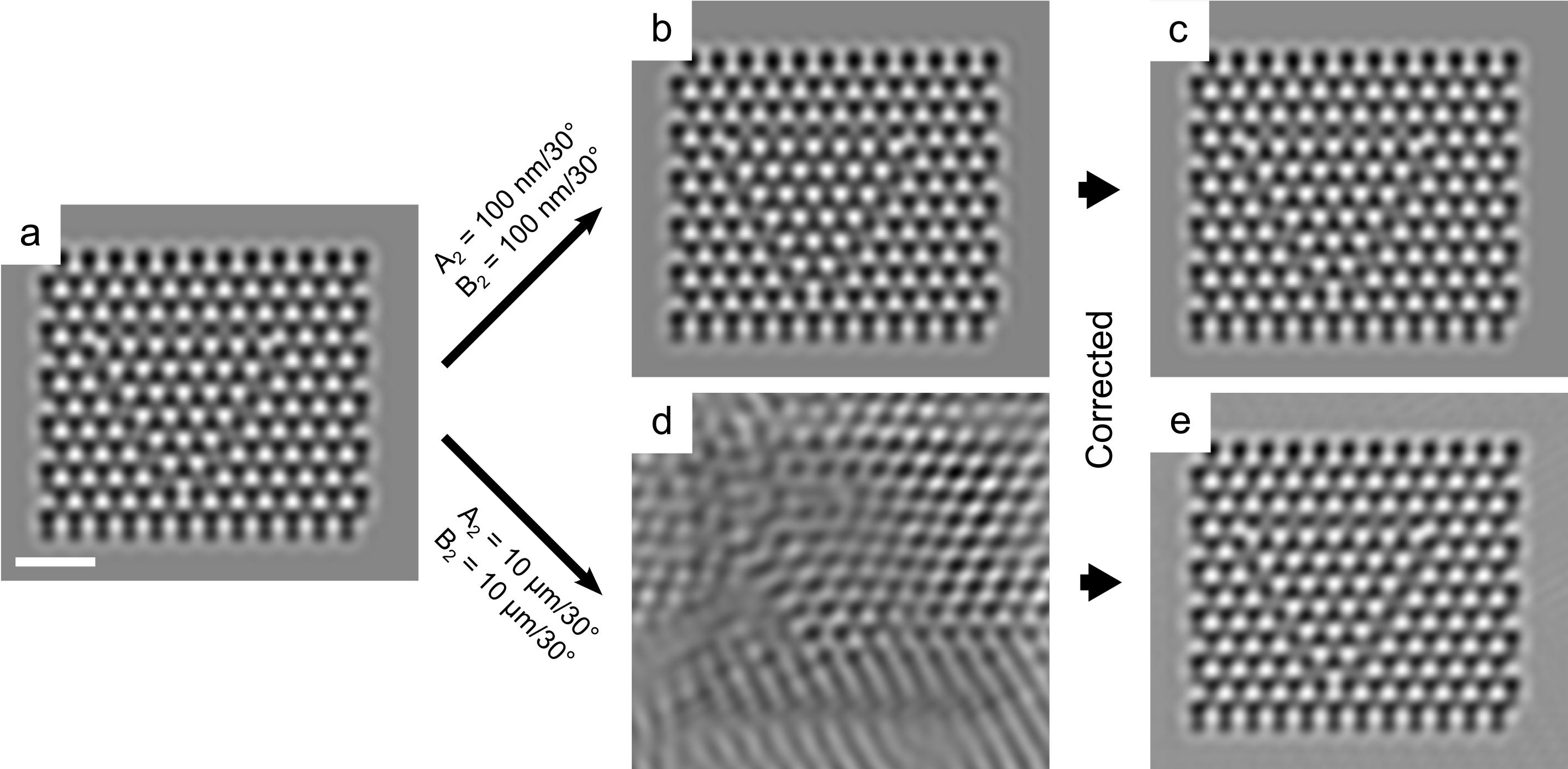}
\caption{{\bf Simulated verification of the correction procedure.} {\bf a}: A simulated reference HRTEM image of MoSe$_2$ flake with an embedded mirror domain with zero anti-symmetric aberrations. The same structure simulated with $A_2 = B_2 = 100$~nm/30$^\circ$ ({\bf b}) and $A_2 = B_2 = 10$~$\mu$m/30$^\circ$ ({\bf d}). 
{\bf c} and {\bf e}: The images of {\bf b} and {\bf d} after applying the correction, with the known parameters. The aberration-free appearance of the image was restored by the method in both cases. The scale bar is 1~nm.}
\label{simulated_correction}
\end{figure}

To verify our approach in the case of the low-order $A_2$ and $B_2$, we first conducted image simulations, where one has precise control over the aberrations influencing the image. In Figure~\ref{simulated_correction}a, a MoSe$_2$ target with an embedded triangular mirror-domain is simulated with $A_2 = B_2 = 0$. The domain-boundaries are visible, and the lattice inside and outside looks identical except for the reversed order of the darker and brighter appearing sublattices. In order to emulate the effect observed in the experimental image of~\ref{residualA2}, a HRTEM image was simulated with $A_2 = B_2 = 100$~nm, and 30$^\circ$ as the azimuthal orientation of both of the aberrations (Figure~\ref{simulated_correction}b). With these parameters the lattice appears completely different inside and outside the mirror-domain. This image was then processed according to equation~\ref{corrector} using the known $A_2$ and $B_2$ coefficients and orientations, and the resulting image is shown in panel~c of Figure~\ref{simulated_correction}. The aberration-free appearance of the image is restored by the correction procedure, as predicted.

We also tested the procedure in more extreme conditions, that is, with $A_2=B_2=10$ $\mu$m. This does not represent typical experimental conditions, but should be considered as an artificial test with very strong aberrations. The simulated image is shown in Figure~\ref{simulated_correction}d, where the whole image is scrambled by the aberrations and interpretation of the image is practically impossible. However, after applying equation~\ref{corrector} with the known parameters, the image is restored to an aberration-free state. Here, it should be pointed out that close inspection of the corrected image reveals some features which are absent in the reference image (panel a), {\em e.g.}, in the vacuum area. This is due to some information being displaced outside the image frame by the (in this case) large point spread function / image delocalization induced by the large aberrations, which then is missing also in the corrected image. Generally speaking, one should avoid interpreting the corrected image within an margin at the image edges with a width determined by the point spread function / image delocalization.

\begin{figure}
\includegraphics[width=.8\textwidth]{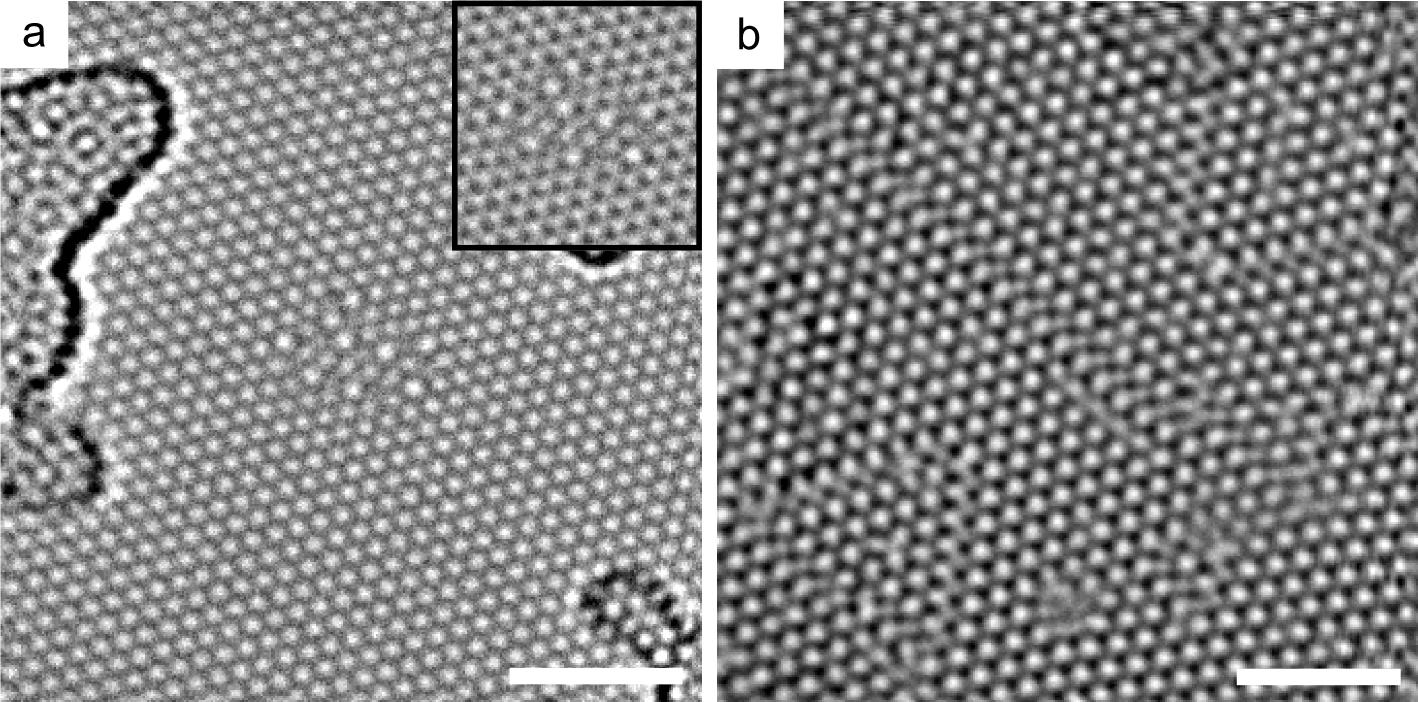}
\caption{{\bf Experimental verification of the correction method.} {\bf a}: The AC-HRTEM image of Figure~\ref{residualA2}a after application of the correction method with $A_2=150$~nm/17$^\circ$. The hexagonal appearance of the graphene lattice is restored and the structure of the defect is easier to analyze. {\bf b}: The AC-HRTEM image of Figure~\ref{residualA2}b after the correction method was applied with $A_2=75$~nm/15$^\circ$. The MoSe$_2$ lattice has identical appearance on both sides of the mirror-twin-boundary as expected for an aberration-free image. The scale bars are 2~nm.}
\label{corrected_exps}
\end{figure}

In order to verify our method also in an experimental situation, we took the AC-HRTEM images of figure~\ref{residualA2}, and applied equation~\ref{corrector}. In the case of graphene, we iteratively found the amplitude and orientation of $A_2=150$~nm/17$^\circ$ which resulted in the removal of visible asymmetry between the two sublattices of graphene. The corrected experimental image is presented in Figure~\ref{corrected_exps}a, with the original image shown as inset. The graphene lattice has now the expected hexagonal appearance, and interpretation of the image is made more straight-forward. For the case of MoSe$_2$, $A_2=75$~nm/15$^\circ$ was found to restore the expected appearance of the lattice on both sides of the mirror-twin-boundaries in the experimental image. 

An interesting observation in the case of MoSe$_2$ is that the point defects are actually more visible in one of the domains in the original (non-corrected) image, and also the boundaries are easier to locate due to the different appearances of the lattice on the opposite sides of the boundaries. This bears resemblance to the well known effect of moving away from the 'optimal' Scherzer focus in order to make point defects more visible to the eye (see, {\em e.g.} Refs.~\cite{meyer2011,lehtinen2014}). No new information is added to the image by the anti-symmetric aberrations, but features are made more evident to the eye. A clear advantage of this approach is that the aberrations can subsequently be corrected with no information losses, as the anti-symmetric aberrations introduce only phase shifts and the phase contrast transfer function is not altered. If the presence of anti-symmetric residual aberrations is deemed beneficial in some situation, our method can just as well be used to {\em increase} the effect of these aberrations during post-processing.

\begin{figure}
\includegraphics[width=.8\textwidth]{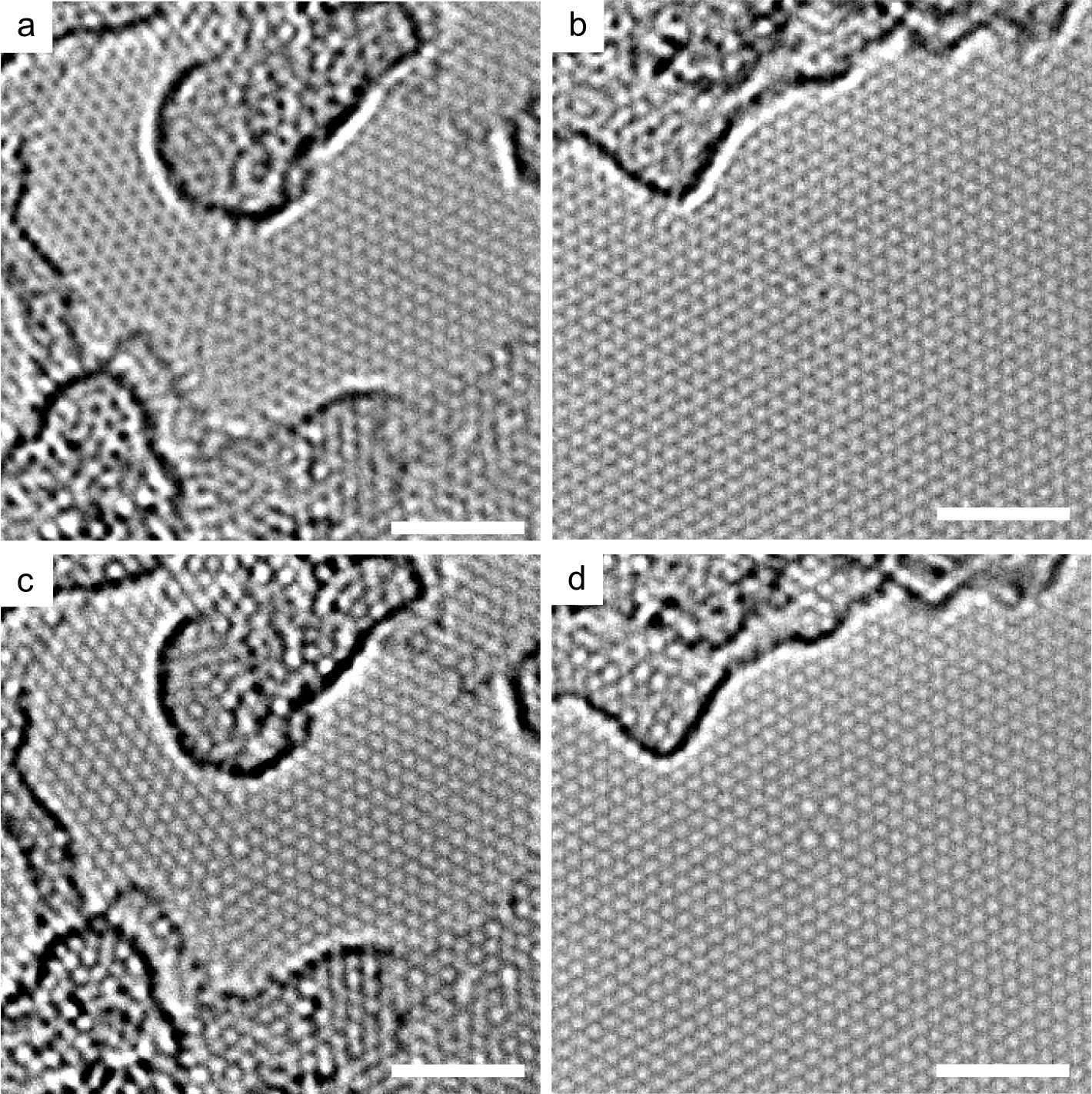}
\caption{{\bf An experimental test with an intentionally introduced large $A_2$ aberration.} {\bf a} and {\bf b}: AC-HRTEM images of a tilt grain boundary and a point defect in graphene with $A_2$ set to $1.0\pm0.1$~$\mu$m/48$^\circ$ (as measured by the corrector software). {\bf c} and {\bf d}: The previous two frames after correcting for 0.9~$\mu$m/42.5$^\circ$ of $A_2$ and 1.0~$\mu$m/41.5$^\circ$ of $A_2$, respectively. In both cases images free of $A_2$ are recovered and the imaged structure is easier to analyze. The scale bars are 2~nm.}
\label{corrected_exps2}
\end{figure}

Another experimental test was conducted by willfully introducing strong $A_2$ with the corrector of the microscope and acquiring images with such an ill-tuned system ($A_2 = 1.0\pm0.1$~$\mu$m/48$^\circ$, as measured by the corrector software). In Figure~\ref{corrected_exps2} a and b, two locations in the graphene sample are shown. In panel a it is clear that a tilt grain boundary runs through the center of the image, separating two areas with different lattice orientations. However, the structure of the boundary cannot be analyzed in the image, and the graphene lattice has a completely different appearance on the opposite sides of the boundary (due to different relative orientations of the lattices relative to the $A_2$ orientation). In panel b, a point defect in the graphene lattice can be detected, but again its structure remains hidden by $A_2$. The correction method was applied to both images using the $A_2$ value of 1~$\mu$m at an angle of 41.5$^\circ$ (measured by the corrector software) as the starting point, and fine tuning the parameters based on visual inspection of the image. In both cases the effect of $A_2$ was remedied. In the case of the grain boundary (panel c) a 5-7 dislocation is now visible, as expected for a tilt-grain-boundary in graphene~\cite{grantab2010anomalous,yazyev2010topological,huang2011,kurasch2012}. The image of the point defect (panel d) shows the recongnizable pattern of a reconstructed divacancy in graphene~\cite{banhart2010structural} after correction. 

\section{Conclusion}

To conclude, we have presented a method for correcting the anti-symmetric sub-group of aberrations in HRTEM images during numerical post-processing in the case of weakly scattering objects, and verified its applicability in the case of the often dominant low-order $A_2$ and $B_2$ aberrations through simulations and experiments with graphene and 2D MoSe$_2$. This procedure can be performed on a single conventional HRTEM image, that is, retrieval of the object wave is not necessary. The requirement of a weakly scattering object imposes a strong limitation on the range of materials where the method is applicable. However, the intensively studied class of 2D materials, with graphene as the prime example, do satisfy this condition at commonly used acceleration voltages such as 80~kV~\cite{Lee201239}. The contribution of the anti-symmetric aberrations is removed by applying the same aberrations with the opposite phase to the Fourier transform of the recorded intensity image and subsequently inverting the Fourier transform, that is, $i_{c}({\bf x}) = |\mathcal{F}^{-1}\{ e^{i\chi_{as}({\bf u})}\cdot \mathcal{F}\{i_z({\bf x})\}\}|$. We have presented the theoretical justification of the approach. The method was demonstrated on simulated and experimental HRTEM images suffering from residual anti-symmetric aberrations ($A_2$ and $B_2$). By applying the method, images with strongly reduced aberrations could be produced. In fact, the anti-symmetric aberration coefficients can be freely adjusted {\em post-situ} using the presented method.

A prerequisite for applying this method is prior knowledge of the aberration coefficients. In the presented examples, the values measured by the corrector software were used, or alternatively the coefficients were found by narrowing down the coefficients leading to the correct appearance of the graphene and MoSe$_2$ lattices through trial and error. The latter approach has an interesting implication: this method can actually be used to estimate the aberration coefficients when there is prior knowledge on how an aberration-free image should look like. Great care should be taken, however, as the problem can be underdetermined, and multiple aberration coefficient combinations can lead to the same final result.

The method allows some flexibility during the acquisition of images: As the anti-symmetric residual aberrations can be corrected during post-processing, it is not imperative to have the corrector 'perfectly' tuned at all times. A reasonably good state of correction is important during acquisition, however, in order to enable the operator of the microscope to recognize the imaged features (compare, {\em e.g.}, Figure~\ref{corrected_exps2} b and d). The loosened requirements for the tuning of the anti-symmetric aberration coefficients can accelerate the corrector tuning procedure. As the time available for the tuning is limited due to, {\em e.g.}, gradual drift of the corrector state~\cite{Barthel20136}, an improved final state of correction can be achieved by the faster tuning procedure.

\vspace{.5cm}
\noindent{\bf Acknowledgements}

\noindent We are very grateful to our group member, senior professor Harald H. Rose for invaluable advice during this work. We would also like to thank Hannu-Pekka Komsa, Arkady V. Krasheninnikov and Martin Linck for helpful discussions, and Nilesh Vats for sample preparation. O.L, D.G., Z.L. and U.K. gratefully acknowledge the funding by the DFG (German Research Foundation) and the Ministry of Science, Research and the Arts (MWK) of Baden-Wuerttemberg in the framework of the SALVE (Sub Angstrom Low-Voltage Electron Microscopy) project. O.L. acknowledges support from the Finnish Cultural Foundation. M.B.W., M.-W.C. and A.K. acknowledge funding from European Research Council (grant no. 240076, FLATRONICS: Electronic devices based on nanolayers)

\end{document}